\documentclass[prl,aps,twocolumn,showpacs,floatfix]{revtex4}
\usepackage{epsfig}
\usepackage{amssymb}
\usepackage{amsmath}
\usepackage{amsfonts}
\usepackage{epstopdf}
\usepackage{graphicx,color}

\newcommand {\be} {\begin{equation}}

\newcommand {\ee} {\end{equation}}

\begin{document}

\author{N.~Teneh$^{a}$, A.~Yu.~Kuntsevich$^{b}$, V.~M.~Pudalov$^b$, T.~M.~Klapwijk$^{c}$, M.~Reznikov$^{a}$}
\address{$^{a}$ Solid state institute, Technion, Haifa, 3200 Israel\\
$^{b}$
LPI, 119991 Moscow, Russia \\
$^{c}$
Delft University of Technology, 2628 CJ
Delft, The Netherlands}

\title{Thermodynamic magnetization of a strongly interacting two-dimensional system }

\begin{abstract}

We report thermodynamic magnetization measurements of  a 2-dimensional electron
gas for several high mobility Si-MOSFETs. The low-temperature magnetization
is shown to be strongly sub-linear function of the magnetic field. The
susceptibility determined from the zero-field slope diverges as
$1/T^{\alpha}$, with $\alpha=2.4\pm0.2$ even at high electron densities, in apparent
contradiction with the Fermi-liquid picture.
\end{abstract}


\date{\today}

\maketitle Magnetic properties of the strongly-interacting electron gas have
long been a subject of intensive theoretical and experimental investigations.
The Coulomb interaction,  as reflected by the exchange term, favors parallel
spins, and therefore leads to ferromagnetism, whenever it is strong enough. The
strength of the interaction is determined by the ratio between the typical
Coulomb and kinetic energies, and is customarily characterized by a
dimensionless parameter $r_s=\rho/a_B$, $a_B$ being the Bohr radius in the
material, and $\rho \sim n^{-1/d}$ is the effective distance between electrons
in $d$ dimensions; in two dimensions  $\rho=1/\sqrt{\pi n}$, $n$ being the
electron density. When $n$ decreases, the relative effect of the interactions
becomes stronger, contrary to naive expectations. At zero temperature and
sufficiently low density the clean single-valley system of itinerant electrons
was predicted to become ferromagnetic, a phenomenon called Stoner
instability\,\cite{Stoner}. In  two dimensions, according to the Mermin-Wagner
theorem, finite temperature destroys the ferromagnetic order.

In a real system there is always some degree of disorder, which favors an
antiferromagnetic interaction. Indeed, the ground state of two localized
spins is a singlet, much like a hydrogen molecule. A system of many localized
spins preserves the tendency to order neighboring spins in opposite directions,
that is, the coupling is antiferromagnetic;
see\,\cite{BhatReview} for a review. Intensive investigations of magnetic
properties of doped semiconductors, particularly phosphorus doped Si, in the 1980's
led to a substantial understanding of the interplay
between interactions and disorder.  The experimentally observed divergence of the
susceptibility as $1/T^{\alpha}$, with $
\alpha\approx 0.6$, on the low-density side of the
metal-insulator transition has been well-understood\,\cite{BhatReview}; however the persistence of the
divergence on the high-density side remains a puzzle.

Two-dimensional gated structures like MOSFETs allow for a gradual change in
electron density, and hence interaction strength, without strongly affecting
sample disorder. Interest in the magnetic properties of two-dimensional
electron gas (2DEG) was stimulated by the observation of strong suppression of
conductivity in Si MOSFETs by an in-plane magnetic field\,\cite{simonyan},
which polarizes electrons and drives the system into an insulating phase.
Scaling analysis of the magnetoresistance led the authors of\,\cite{SKDK,VZ2} to
suggest a quantum phase transition into a ferromagnetic state at the density of
the metal-insulator transition, $n_c$\,\cite{comment}. This claim was contested
in\,\cite{Prus2003}  and \,\cite{PGK2001} on the basis of thermodynamic and
Shubnikov-de Haas measurements, respectively.

In this Letter we use the technique developed and described in\,\cite{Prus2003}
to study the magnetization at elevated temperatures. This technique measures
the recharging of the gate-to-2DEG capacitor due to a change in the 2DEG
chemical potential $\mu$. By modulating an external magnetic field with the
amplitude $\delta B(\omega)$, while keeping the gate voltage constant we are
able to measure a gate charge modulation $\delta Q(\omega)$ given by
\begin{equation}\label{dmudb}
\delta Q(\omega) = \frac {C(\omega)}{e} {\frac{\partial\mu}{\partial B}}\;\delta
B(\omega),
\end{equation}
%
%
where the capacitance $C(\omega)$ is measured independently by modulating the
gate voltage at the same frequency.  We note that all the quantities in Eq.
\ref{dmudb} also depend on the electron density $n$; this dependence is not
explicitly indicated. By virtue of the Maxwell relation $\partial\mu/\partial
B=-\partial m /\partial n$, $\partial\mu/\partial B$ can be  expressed as the
derivative of the magnetization per unit area $m$ with respect to the density.
In principle, integrating $\partial m /\partial n$ over the density from $n=0$
would give $m(n)$.

In this Letter we present measurements from deep into the insulating region to
far into the metallic phase, allowing us to integrate $\partial m /\partial n$
from almost $n=0$. The magnetization $m(n)$ thus obtained exhibits a
surprisingly strong temperature dependence at low magnetic field $\mu_B B\ll T$
even at high densities. This result contradicts expectations from the Fermi
liquid theory. Indeed, at high densities interactions are relatively weak,
therefore magnetization should be temperature independent  at temperatures small
compared to the Fermi energy $E_F$.

Extension of the measurements deep into the insulating regime became possible
due the lower sample resistance at elevated temperatures, together with good
quality of the contacts. It is important to note that the
expression (\ref{dmudb}) holds even if the capacitance drops and acquires
imaginary part due to the contact and channel resistance, which happens at the
resistance about 500~MOhms. This facilitates measurements deep into insulating
region down to less the half the density of the metal-insulator transition in
the sample.

\begin{figure}
\vskip.05in
\begin{center}
\centerline{\psfig{figure=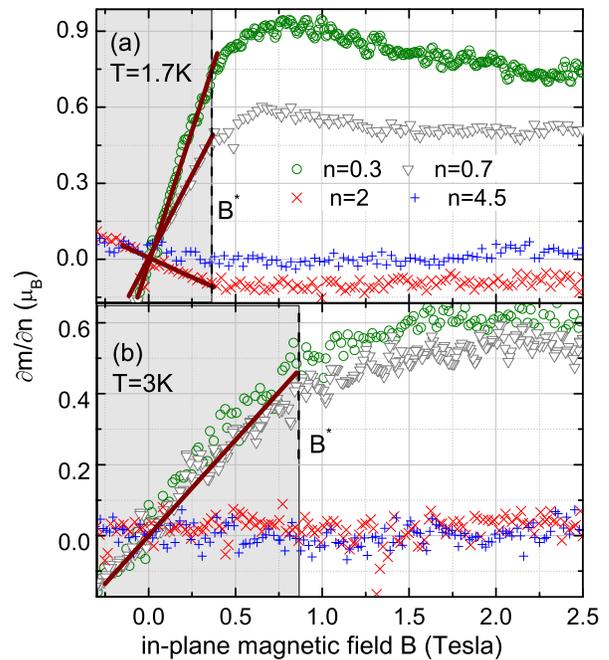, width=230pt}} \caption{Differential
magnetization, $\partial m/\partial n $, for several densities $n$ are given in
$10^{11}{\rm cm^{-2}}$. Gray area mark the linear regime
$B<B^*$.}\label{baredata}
\end{center}\vskip -0.5in
\end{figure}

Our measurements were performed on several high-mobility Hall bar shaped Si-MOS
structures with 2DEG located on the (100) interface between Si surface and
SiO$_2$. Such a 2DEG possesses a two-fold valley degeneracy in addition to its
spin degeneracy. We used two groups of samples: those fabricated in Russia (R),
similar to those used in Ref.\,\cite{simonyan,Prus2003,pudalovsdh} and those
made in Holland (H), used in Ref.\,\cite{SKDK,VZ2}. The in-plane magnetic field
was modulated at frequency $\omega/2\pi=6.1$~Hz with amplitude $40$~mT.
Standard He4 pumping and heating were used to set the temperature in the range
1.7-13K. We present the results obtained from the R sample with peak mobility
3.4${\rm m^2/Vs}$ and H sample with 3.3${\rm m^2/Vs}$ at 1.7K, measured most
extensively; the data for several other R and H samples were very similar. An
example of the data collected at different temperatures and electron densities
is shown in Fig.\,\ref{baredata}. At low field, $\partial m/\partial n (B)$
grows linearly below some temperature-dependent field $B^*$ indicated by the
gray area in Fig.\,\ref{baredata}, above which the slope changes. The maximal
$\partial m/\partial n (B)$ at lowest achievable density and temperature
reaches $0.9\mu_B$ at $B\approx 0.3$~T, indicating almost full spin
polarization.

\begin{figure}
\begin{center}
\centerline{\psfig{figure=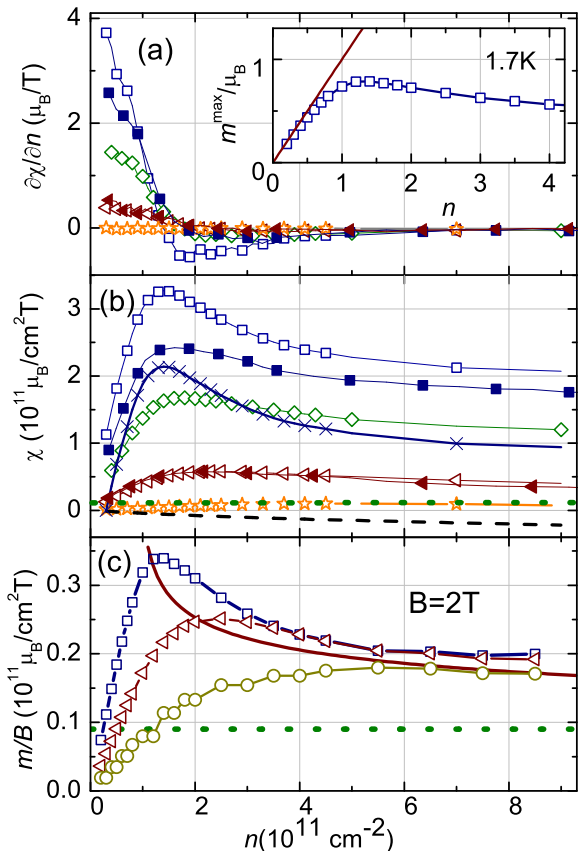,width=230pt}} \caption{ (a) Derivative of
the low-field susceptibility with respect to the density $\partial\chi/\partial
n$ for different temperatures. Hollow symbols-R sample, bold symbols - H
sample.  (b)~$\chi$ obtained by integration; Dotted line (also in c) - non-renormalized
Pauli susceptibility, dashed line--subtracted diamagnetic contribution.
$\times$ - the same as $\square$ integrated from the density $n_L$. (c)
Susceptibility defined as $m/B$ at 2T. Solid line - the Shubnikov-de Haas
susceptibility\,\cite{pudalovsdh}. The temperatures are indicated by the
symbols: $\square$ - 1.7K;
{\large $\diamond$} -2.4K;
{\large $\triangleleft$} - 4K;
{\large $\circ$} - 7K; {\large $\star$} - 13K. Inset-- maximum of the magnetization in the field $0<B<2$T at
$T=1.7$K, $m$ and $n$ are in units of $10^{11}{\rm cm^{-2}}$}.
\label{chi}
\end{center}\vskip -0.5in
\end{figure}

Let us first consider the low-field slope of the data, $\partial^2 m/\partial n
\partial B|_{B=0}=\partial \chi/\partial n|_{B=0}$, plotted in Fig.\,\ref{chi}
as a function of density at different temperatures. The slope is
large and positive at low temperature and density. As the density
increases, the slope decreases, changes sign, and finally almost
tends to zero at highest densities. As a function of temperature
the slope decreases rapidly almost to zero at $T=13K$. There is
some small negative slope left at highest temperatures in the R
samples, which we attribute to the diamagnetic contribution due to
the finite thickness of the 2DEG\,\cite{Prus2003}. This negative
slope is absent for H samples\,\cite{comment1}.

In order to obtain the zero field susceptibility we integrate $\partial \chi/\partial n$:
\be \chi(n,T)=\int\limits_0^n \partial \chi/\partial n(n',T)dn'\label{int}\ee
We were able to measure $\partial \chi/\partial n$ down to densities $0<n_{L}(T)< n_c$,
so we must extrapolate the integrand in (\ref{int}) down to zero density.
For simplicity we took $\partial \chi/\partial n$ to be constant and
equal to its value at $n_L(T)$. In order to get the spin part of
susceptibility we need to subtract the temperature-independent diamagnetic susceptibility. This we did by equating the high temperature susceptibility
to the non-renormalized Pauli one. The results of the integration under
these assumptions are depicted in Fig.\,\ref{chi}b, together with the
subtracted diamagnetic contribution.

Before discussing the results, let us first examine possible sources of
inaccuracy. The diamagnetic contribution is small, and therefore can affect
only the high-temperature susceptibility. The largest error may come from the
extrapolation of $\partial \chi/\partial n$ to zero density, as well from the
uncertainty $\approx10^{10}{\rm cm^{-2}}$ in the density itself\,\cite{n0}. We note that even if
we make the extreme assumption that the magnetization is zero below
$n_L(T)$, thus significantly underestimating the susceptibility, we would get
qualitatively similar results, as also shown in Fig.\,\ref{chi}b.

\begin{figure}[t]
\vskip.05in
\begin{center}
\centerline{\psfig{figure=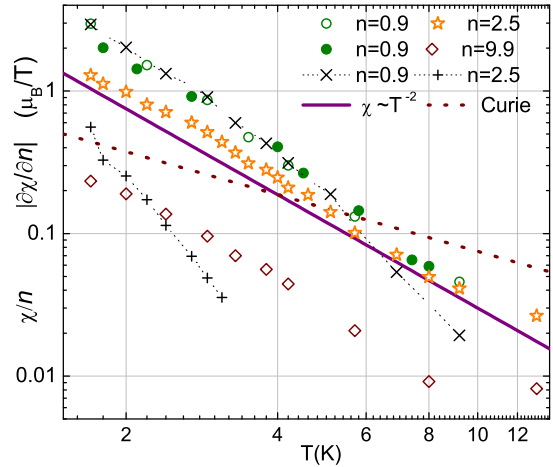,width=230pt}}
\caption{Low-field susceptibility $\chi$ per electron vs. temperature for different densities,
indicated in units of $10^{11}{\rm cm^{-2}}$. Hollow symbols-R sample, bold symbols - H
sample. Also $|\partial\chi/\partial
n|$ is shown for comparison with $\times$ and +.
Dashed line line-$\chi\propto 1/T^2$, solid line-Curie susceptibility.
}\label{chi(T)}
\end{center}\vskip -0.5in
\end{figure}

The most striking features of the susceptibility $\chi$ are (i) its
low-temperature value, at maxima exceeding the Pauli one by the factor of 40,
and (ii) strong temperature and relatively weak density dependence at high
densities. The temperature dependence of the susceptibility per electron for
several densities is shown in Fig.\,\ref{chi(T)}. It can be fit reasonably
well with a power law $\chi\propto 1/T^{\alpha}$, with $\alpha\approx 2.4\pm 0.2$.
Not only does the susceptibility diverge faster than the independent-spin Curie
susceptibility, $\mu_B^2/T$, but its low-temperature low-density value exceeds the Curie
value, indicated by the solid line in Fig.\,\ref{chi(T)}. We emphasize that neither the
susceptibility value, nor its temperature dependence change qualitatively when
the system passes through the metal-insulator transition at $n_c\approx
8.5\cdot 10^{10}{\rm cm^{-2}}$. Remarkably, the results for the R and H
samples, shown in Figs.\,\ref{chi},\ref{chi(T)} by the bold and hollow symbols
respectively, are very similar. The largest  contribution to the integral (\ref{int}), which determines $\chi$, even at
high densities comes from $\partial\chi/\partial n$ at low densities. Therefore it is important to stress that $\partial\chi/\partial
n$ also diverges as $1/T^{\alpha}$, even deep in the metallic regime, as indicated in Fig.\,\ref{chi(T)}.

\begin{figure}[t]
\begin{center}
\centerline{\psfig{figure=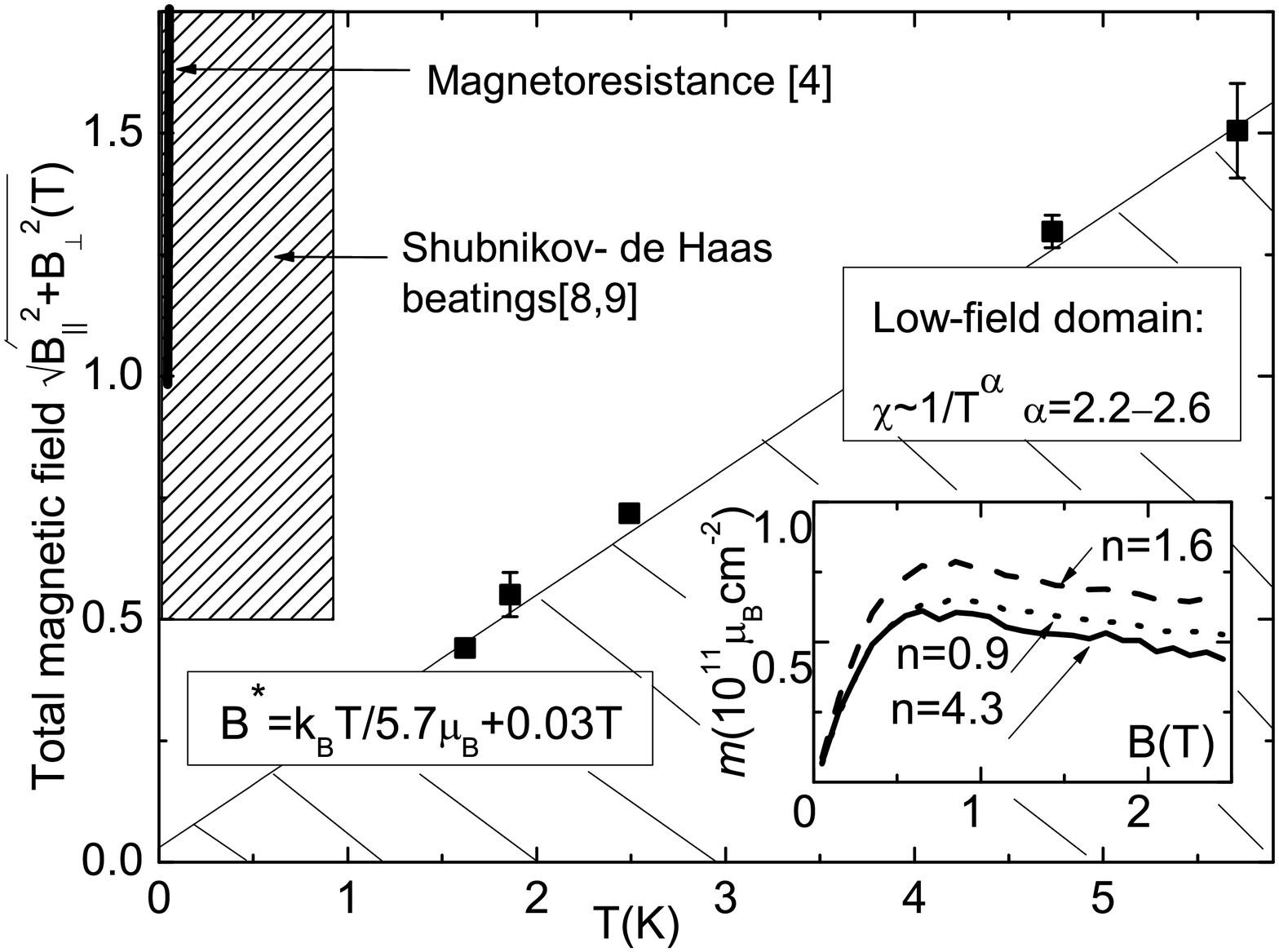,width=250pt}} \caption{Comparison with
transport measurements. $\blacksquare$ marks almost density-independent $B^*$. Arrows
indicate the domains of the previous measurements of $\chi$.
Inset: field dependence of magnetization for R sample at 1.7K.}\label{summary}
\end{center}\vskip -0.5in
\end{figure}

The magnetic moment can be obtained similarly to the susceptibility, by
integration of $\partial m(n,T, B)/\partial n$ over $n$ for given values of $B$
and $T$. Since at low-temperature $\partial m/\partial n$ saturates, and even
drops for $B>B^*$, as seen in Fig.\,\ref{baredata}a, $m(B)$ behaves similarly;
see the inset in Fig.\,\ref{summary}. This saturation resolves the apparent
contradiction between the current thermodynamic and
earlier\,\cite{SKDK,pudalovsdh} transport susceptibility measurements. Indeed,
all the transport measurements to date were performed in the high-field domain,
when the Zeeman energy is larger than the temperature, whereas we observe the
divergent susceptibility solely in the low field domain. As seen in
Fig.\,\ref{summary} $B^*$ depends linearly on temperature. Additionally,
we note that $B^*$ appears to be density-independent for $n\lessapprox 3\cdot
10^{11}{\rm cm^{-2}}$. The low-field domain, at which the zero-field
susceptibility can be determined, lies in the bottom-right corner. In contrast,
both Shubnikov-de Haas, and magnetoresistance data belong to the top left
corner. If one takes the magnetic moment at e.g. 2T, which is the typical {\it
total} field in\,\cite{pudalovsdh,SKDK,VZ2}, he would get the value of the
susceptibility, defined as $\chi(B)=m(B)/B$, similar to obtained
in\,\cite{SKDK,pudalovsdh}, as shown in Fig.\,\ref{chi}c.

In our previous work\,\cite{Prus2003} we could only measure $\partial m
/\partial n$ down to $\approx n_c$, and therefore we chose to integrate it from
a high density at which we assumed $m$ to be known and temperature independent.
We used the susceptibility obtained from the Shubnikov-de Haas measurements in
the high-field domain, (similar assumption was later used in\,\cite{shashkintd})
as the initial value for the integration in both high and low field domains. As
the present work shows the temperature dependence of the susceptibility in the
low field domain cannot be ignored.

A susceptibility diverging faster than $1/T$ indicates a ferromagnetic
interaction between spins. If, however, the low temperature extrapolation of
$B^*=k_BT/5.7\mu_B+0.03T$ in Fig.\,\ref{summary} is taken seriously, it means that the divergence should be cut off at about 100mK.
A divergent susceptibility contradicts numerical calculations\,\cite{MarchiFleri}
for a two-valley system, which predicts no tendency to ferromagnetism. At high
enough density one would expect the susceptibility to be determined by
excitations in the vicinity of the Fermi level. Interaction corrections to
susceptibility in a clean system were calculated in\,\cite{Fin, Maslov}. Although
the results differ in prefactors, both predict a correction to the
susceptibility of the order of $T/E_F$. It is important to stress that not only does
$\chi$ diverge as $1/T^{\alpha}$, but so does $\partial\chi/\partial n$, even
deep in the metallic regime at densities $n=2-4\cdot 10^{11}{\rm cm^{-2}}$, in
clear contradiction with\,\cite{Fin, Maslov}. We therefore think the divergent
susceptibility is disorder-related and signals the presence of local moments in
the 2D system  up to high densities; numerical calculations indeed predict
susceptibility enhancement by disorder\,\cite{DePalo}. This suggestion does not
contradict numerous existing transport data, since localized states are very
rarely seen in transport, which probe the electrons on the time scale of ps. In
contrast, thermodynamic measurements probe all the states which can be
recharged on the time scale set by the magnetic field modulation. On the other
hand, it is important to emphasize that the maximal magnetization of the
system, shown in the inset in Fig.\,\ref{chi}a, is very close to $n\mu_B$ at low
densities. This rules out the contribution of distant magnetic centers,
existing {\it in addition} to the 2DEG (such as, e.g., localized states in the
bulk Si, or deep levels at the interface).


A clue about the origin of the anomalous susceptibility may come from the
magnetization drop at low temperatures in the intermediate magnetic field, above $B^*$ and below
some 2.5T, visible in the inset in Fig.\,\ref{summary}; the
magnetization grows again at even higher fields. For a magnetic field coupled only to
the spins, the magnetization should grow with the field. Therefore this
drop must be related either to the orbital effects, or to the field induced
variation in the number of electrons contributing to the signal.


In conclusion, we observed a strongly enhanced divergent susceptibility at low temperatures in
high-mobility MOSFET's. We find no qualitative change of the susceptibility behavior in the
vicinity of the metal-insulator transition density $n_c$, and the divergence persists deep into the
high density metallic phase, similar to the observations on phosphorus doped Si.

We thank A.~Finkel'stein for motivating us to recheck susceptibility at
elevated temperatures, and D.Maslov and I.Burmistrov for discussions. This
work was supported by the Israeli National Science foundation, RFBR, Programs
of RAS, Russian Ministry for Education and Science, and the Program "Leading
scientific schools".


\begin{thebibliography}{qq}
\bibitem{comment} $n_c$ is the density below which the metalic temperature dependence of
    resistivity switches to the insulating one.
\bibitem{comment1} p-type substrate dopping estimated from C-V measurements is $2\cdot 10^{15}{\rm
    cm^{-3}}$ for R  samples,  and $3\cdot 10^{16}{\rm   cm^{-3}}$ for the H ones. Higher dopping
    means thinner 2-D gas, and therefore smaller diamagnetic contribution. Indeed,
    in\,\cite{shashkintd} diamagnetic contibution was not observed.
\bibitem{n0} Since the magnetic field was in the sample plain, we could not
    measure the Hall simultaneously, and zero density was determined from
    comparison between the Hall and capacitance measurements in a different
    run.
\end{thebibliography}
\end{document}